\documentclass[useAMS,usenatbib]{mn2e}
\usepackage{times}
\usepackage{graphicx}
\usepackage{subfigure}
\usepackage{psfrag}
\usepackage{amsmath}
\usepackage{amssymb}

\def\reff@jnl#1{{\rm#1\/}}
\def\aj{\reff@jnl{AJ}}                 % Astronomical Journal
\def\araa{\reff@jnl{ARA\&A}}           % Annual Review of Astron and Astrophys
\def\apj{\reff@jnl{ApJ}}               % Astrophysical Journal
\def\apjl{\reff@jnl{ApJ}}              % Astrophysical Journal, Letters
\def\apjs{\reff@jnl{ApJS}}             % Astrophysical Journal, Supplement
\def\ao{\reff@jnl{Appl.Optics}}        % Applied Optics
\def\apss{\reff@jnl{Ap\&SS}}           % Astrophysics and Space Science
\def\aap{\reff@jnl{A\&A}}              % Astronomy and Astrophysics
\def\aapr{\reff@jnl{A\&A~Rev.}}        % Astronomy and Astrophysics Reviews
\def\aaps{\reff@jnl{A\&AS}}            % Astronomy and Astrophysics, Supplement
\def\azh{\reff@jnl{AZh}}               % Astronomicheskii Zhurnal
\def\baas{\reff@jnl{BAAS}}             % Bulletin of the AAS
\def\jcap{\reff@jnl{JCAP}}             % Journal of Cosmology and Astroparticle Physics
\def\jrasc{\reff@jnl{JRASC}}           % Journal of the RAS of Canada
\def\memras{\reff@jnl{MmRAS}}          % Memoirs of the RAS
\def\mnras{\reff@jnl{MNRAS}}           % Monthly Notices of the RAS
\def\pra{\reff@jnl{Phys.Rev.A}}        % Physical Review A: General Physics
\def\prb{\reff@jnl{Phys.Rev.B}}        % Physical Review B: Solid State
\def\prc{\reff@jnl{Phys.Rev.C}}        % Physical Review C
\def\prd{\reff@jnl{Phys.Rev.D}}        % Physical Review D
\def\prl{\reff@jnl{Phys.Rev.Lett}}     % Physical Review Letters
\def\pasp{\reff@jnl{PASP}}             % Publications of the ASP
\def\pasj{\reff@jnl{PASJ}}             % Publications of the ASJ
\def\qjras{\reff@jnl{QJRAS}}           % Quarterly Journal of the RAS
\def\skytel{\reff@jnl{S\&T}}           % Sky and Telescope
\def\solphys{\reff@jnl{Solar~Phys.}}   % Solar Physics
\def\sovast{\reff@jnl{Soviet~Ast.}}    % Soviet Astronomy
\def\ssr{\reff@jnl{Space~Sci.Rev.}}    % Space Science Reviews
\def\zap{\reff@jnl{ZAp}}               % Zeitschrift fuer Astrophysik
\def\nat{\reff@jnl{Nature}}            % Nature
\newcommand{\Trise}{T_{\rm rise}} 
\newcommand{\Tfall}{T_{\rm fall}} 

\title[Automated photometric classification of supernovae]
{A simple and robust method for automated photometric
  classification of supernovae using neural networks}
\author[N.V.~Karpenka, F.~Feroz and M.P.~Hobson]  
{N.V.~ Karpenka$ ^{1} $\thanks{E-mail: nkarp@fysik.su.se}, F.~Feroz$ ^2 $ 
and M.P.~Hobson$ ^2 $\\ 
$^1$The Oskar Klein Centre for Cosmoparticle Physics, Department of Physics, Stockholm University, AlbaNova, SE-106 91 Stockholm, Sweden\\
$^2$Astrophysics Group, Cavendish Laboratory, J.J.~Thomson Avenue, Cambridge CB3 0HE, UK\\}

\date{Accepted 2012 November 14. Received 2012 October 11; in
  original form 2012 August 7}

\pagerange{\pageref{firstpage}--\pageref{lastpage}}
\pubyear{2012}

\begin{document}
\label{firstpage}
\maketitle
\begin{abstract}
A method is presented for automated photometric classification of
supernovae (SNe) as Type-Ia or non-Ia. A two-step approach is adopted
in which: (i) the SN lightcurve flux measurements in each observing
filter are fitted separately to an analytical parameterised function
that is sufficiently flexible to accommodate vitrually all types of
SNe; and (ii) the fitted function parameters and their associated
uncertainties, along with the number of flux measurements, the
maximum-likelihood value of the fit and Bayesian evidence for the
model, are used as the input feature vector to a classification neural
network (NN) that outputs the probability that the SN under consideration
is of Type-Ia.  The method is trained and tested using data released
following the SuperNova Photometric Classification Challenge (SNPCC),
consisting of lightcurves for 21,319 SNe in total. We consider several
random divisions of the data into training and testing sets: for
instance, for our sample ${\cal D}_1$ (${\cal D}_4$), a total of 10
(40) per cent of the data are involved in training the algorithm and
the remainder used for blind testing of the resulting classifier; we
make no selection cuts. Assigning a canonical threshold probability of
$p_{\rm th}=0.5$ on the network output to class a SN as Type-Ia, for
the sample ${\cal D}_1$ (${\cal D}_4$) we obtain a completeness of
0.78 (0.82), purity of 0.77 (0.82), and SNPCC figure-of-merit of 0.41
(0.50).  Including the SN host-galaxy redshift and its uncertainty as
additional inputs to the classification network results in a modest
5--10 per cent increase in these values. We find that the quality of
the classification does not vary significantly with SN
redshift. Moreover, our probabilistic classification method allows one
to calculate the expected completeness, purity and figure-of-merit (or
other measures of classification quality) as a function of the
threshold probability $p_{\rm th}$, without knowing the true classes
of the SNe in the testing sample, as is the case in the classification
of real SNe data. The method may thus be improved further by
optimising $p_{\rm th}$ and can easily be extended to divide non-Ia
SNe into their different classes.
\end{abstract}

\begin{keywords}
methods: data analysis -- methods: statistical --
 supernovae: general
\end{keywords}

%%%%%%%%%%%%%%%%%%%%%%%%%%%%%%%%%%%%%%%%%%%%%%%%%%%%%%%%%
\section{Introduction}\label{sec:intro}
%%%%%%%%%%%%%%%%%%%%%%%%%%%%%%%%%%%%%%%%%%%%%%%%%%%%%%%%%
Much interest in supernovae (SNe) over the last decade has been
focused on Type-Ia (SNIa) for their use as `standardizable' candles
in constraining cosmological models.  Indeed, observations of SNIa led
to the discovery of the accelerated expansion of the universe
(\citealt{1998AJ....116.1009R}; \citealt{1999ApJ...517..565P}), which
is usually interpreted as evidence for the existence of an exotic dark
energy component.  Ongoing observations of large samples of SNIa
are being used to improve the measurement of luminosity distance as a
function of redshift, and thereby constrain cosmological parameters
further (e.g., \citealt{2009ApJS..185...32K};
\citealt{2012MNRAS.419..513B}; \citealt{2011ApJ...737..102S};
\citealt{2011ApJS..192....1C}; \citealt{march11}) and improve our
knowledge of dark energy (e.g., \citealt{2010MNRAS.406.1759M};
\citealt{2011MNRAS.418.1707B}). Moreover, the gravitational lensing of
SNIa by foreground cosmic structure along their lines-of-sight has
been used to constrain cosmological parameters (\citealt{metcalf99},
\citealt{dodelson06}, \citealt{zentner09}) and the properties of the
lensing matter (\citealt{rauch91}, \citealt{metcSilk},
\citealt{jacob07}, \citealt{kronborg10}, \citealt{jonssonGOODS},
\citealt{jonssonSNLS}, \citealt{Karpenka2012}). In addition to their
central role in cosmology, the astrophysics of SNIa is also of
interest in its own right, and much progress has been made in
understanding these objects in recent years
(e.g. \citealt{2000ARA&A..38..191H}).

Other types of SNe are also of cosmological interest.  Type II Plateau
Supernovae (SNII-P), for example, can also be used as distance
indicators, although only for smaller distances and to lower
accuracy than SNIa. Compared to SNIa, however, for which there is
still uncertainty regarding the progenitor system, SNII-P explosions
are better understood. Furthermore, since SNII-P have only been found
in late-type galaxies, biases from environmental effects will most
probably have a smaller effect on distance measurements using
SNII-P. Thus, the differences between the two types of SNe will result
in different systematic effects, allowing SNII-P data to complement
SNIa analyses \citep{2010ApJ...708..661D}.

Although not used directly in cosmology, other classes of SNe are a
potential source of contamination when attempting to compile SNIa
catalogues, most notably SN Ib/c. The consequences of such
contamination have been considered by \cite{2005ApJ...620...12H}. SN
Ib/c are also of considerable astrophysical interest, in particular
the nature of their progenitors \citep{2007PASP..119.1211F}.

%In addition, more recently, other types of of SNe have been
%discovered. (ADD REF)

The next generation of survey telescopes, such as the Dark Energy
Survey (DES; \citealt{2005ASPC..339..152W},
\citealt{2011AAS...21743316A}), the Large Synoptic Survey Telescope
(LSST; \citealt{2002SPIE.4836...10T}, \citealt{2008arXiv0805.2366I}),
and SkyMapper (\citealt{2005AAS...206.1509S}), are expected to observe
lightcurves for many thousands of SNe, far surpassing the resources
available to confirm the type of each of them
spectroscopically. Hence, in order to take advantage of this large
amount of SNe data, it is necessary to develop methods that can
accurately and automatically classify many SNe based only on their
photometric light curves.

In response to this need, many techniques targeted at SNe photometric
classification have been developed, mostly based on some form of
template fitting (\citealt{2002PASP..114..833P};
\citealt{2006AJ....132..756J}; \citealt{2006AJ....131..960S};
\citealt{2007AJ....134.1285P}; \citealt{2007ApJ...659..530K};
\citealt{2007PhRvD..75j3508K}; \citealt{2008AJ....135..348S};
\citealt{2011ApJ...738..162S}; \citealt{2009ApJ...707.1064R};
\citealt{2010ApJ...709.1420G}; \citealt{2010ApJ...723..398F}).  In
such methods, the lightcurves in different filters for the SN under
consideration are compared with those from SNe whose types are well
establised. Usually, composite templates are constructed for each
class, using the observed lightcurves of a number of well-studied,
high signal-to-noise SNe (see \citealt{2002PASP..114..803N}), or
spectral energy distribution models of SNe. Such methods can produce
good results, but the final classification rates are very sensitive to
the characteristics of the templates used.

To address this difficulty, \cite{Newling2011} instead fit a
parametrised function  to the SN lightcurves. These
post-processed data are then used in either a kernel density
estimation method or a `boosting' machine learning algorithm to assign a
probability to each classification output, rather than simply
assigning a specific SN type.  More recently,
\cite{2012MNRAS.419.1121R} and \cite{2012arXiv1201.6676I} have
introduced methods for SN photometric classification that do not rely
on any form of template fitting, but instead employ a mixture of
dimensional reduction of the SNe data coupled with a machine learning
algorithm. \cite{2012MNRAS.419.1121R} proposed a method that uses
semi-supervised learning on a database of SNe: as a first step they use
all of the lightcurves in the database simultaneously to estimate a
low-dimensional representation of each SN, and then they employ a set
of spectroscopically confirmed examples to build a classification
model in this reduced space, which is subsequently used to estimate
the type of each unknown SN. Subsequently, \cite{2012arXiv1201.6676I}
proposed the use of Kernel Principal Component Analysis as a tool to
find a suitable low-dimensional representation of SNe lightcurves. In
constructing this representation, only a spectroscopically confirmed
sample of SNe is used. Each unlabeled lightcurve is then projected
into this space and a $k$-nearest neighbour algorithm performs the
classification.

In this paper, we present a new method for performing SN
classification that also does not rely on template fitting in a
conventional sense, but combines parametrised functional fitting of
the SN lightcurves together with a machine learning algorithm.  Our
method is very straightforward, and might reasonably even be described
as naive, but nonetheless yields accurate and robust classifications.

%For the SNe in the training set,
%we first perform a dimensionality reduction by fitting the lightcurves
%of each SN with a simple, parametrised, analytical form that is
%sufficiently flexible to represent virtually all types of SNe. The
%fitted parameters, together with a few further statistics associated
%with the fit, then constitute the feature vector input to a
%feed-forward classificaton neural network (NN) whose weights are
%optimised over the training set to recover as often as possible the
%known classes of the SNe. The lightcurve fitting and trained neural
%network may then be used to estimate the class of SNe outside the
%training set. We focus here on classifying SNe as simply Ia or non-Ia,
%although our approach may be straightforwardly generalised to divide
%the non-Ia supernovae into their different classes.

The outline of this paper is as follows. In Sec.~\ref{sec:data}, we
describe the data set used for training and testing in our analysis,
and we present a detailed account of our methodology in
Sec.~\ref{sec:method}.  We test the performance of our approach in
Sec.~\ref{sec:res} by applying to the data and present our
results. Finally, we conclude in Sec.~\ref{sec:conc}.

%%%%%%%%%%%%%%%%%%%%%%%%%%%%%%%%%%%%%%%%%%%%%%%%%%
\section{Post-SNPCC simluated data set }\label{sec:data}
%%%%%%%%%%%%%%%%%%%%%%%%%%%%%%%%%%%%%%%%%%%%%%%%%%

The data set we will use for training and testing our classification
algorithm described below is that released by the SuperNova
Photometric Classification Challenge (SNPCC),
%\footnote{Data can be downloaded from \texttt{http://sdssdp62.fnal.gov/sdsssn/SIMGEN_PUBLIC/SIMGEN_PUBLIC_DES.tar.gz}}, 
which consists of simulations of lightcurves in several filters for
21,319 SNe \citep{2010PASP..122.1415K}.  The simulations were made
using the \texttt{SNANA} package \citep{Kessler2009SNANA} according to
DES specifications. We used the updated version of the simulated
data-set (post-SNPCC), which was made public after the challenge
results were released. This updated data-set is quite different from
the one used in the challenge itself, owing to some bug fixes and
other improvements aimed at a more realistic simulation of the data
expected for DES. The data set contains SNe of Types Ia, Ib/c, Ib, Ic,
IIn, II-P and II-L, sampled randomly with proportions given by their
expected rates as a function of redshift. Averaged over all redshifts,
the approximate proportions of each type are 25, 1, 7, 5, 9, 51 and 2
per cent, respectively.

This large data set, which we denote by $\mathcal{D}$, is made up of
two simulated subsamples: a small `spectroscopically-confirmed' sample
of 1,103 SNe, which we denote by $\mathcal{S}$, and a `photometric'
sample of 20,216 SNe, denoted by $\mathcal{P}$. The $\mathcal{S}$
subsample consists of simulated lightcurves for a set of SNe that one
could follow-up with a fixed amount of spectroscopic resources on each
of a 4-m and 8-m class telescope. The magnitude limits were assumed to
be 21.5 ($r$ band) for the 4-m and 23.5 ($i$ band) for the 8-m
telescope. Since spectroscopy is more demanding than photometry,
$\mathcal{S}$ consists of SNe that on average have higher observed
brightnesses and much lower host-galaxy redshifts than those in the
photometric sample $\mathcal{P}$. Consequently, $\mathcal{S}$ is not a
random subset of $\mathcal{D}$, but instead has a much higher fraction
of SNIa.

For each one of the SNe in $\mathcal{D}$, we denote the data by
$\mathbf{D}_i = \{t_{i,k}^{\alpha}, F_{i,k}^{\alpha}, \sigma_{
  i,k}^{\alpha}\}$, where $i$ indexes the SNe, $\alpha \in \{\rm
g,r,i,z\}$ denotes the filter, $k = 1, 2, \cdots, n_{i}^{\alpha}$
indexes the number of flux measurements in filter $\alpha$,
$t_{i,k}^{\alpha}$ is the time for the given measurements,
$F_{i,k}^{\alpha}$ is the \texttt{SNANA} `calibrated' flux measured at
time $t_{i,k}^{\alpha}$ and $\sigma_{i,k}^{\alpha}$ is the
corresponding uncertainty. The lightcurve in each filter for each SN
is measured on an irregular time grid that differs from filter to
filter and between SNe. For each SN, between 16 to 160 measurements
were made (between 4 and 40 measurements in each filter), with a
median value of 101 measurements. We note that some of the SNe
lightcurves were observed only before or after the peak in emission,
but were still included in our analysis.

%%%%%%%%%%%%%%%%%%%%%%%%%%%%%%%%%%%%%%%%%%%%%%%%%%
\section{Analysis methodology}\label{sec:method}
%%%%%%%%%%%%%%%%%%%%%%%%%%%%%%%%%%%%%%%%%%%%%%%%%%

In order to perform the SNe classification, we adopt a two-step
process:
\begin{enumerate}
\item
SNe lightcurves are first fitted to an analytical parameterised
function, in order to standardise the number of input parameters for
each SNe that are used in NN training. Provided that the function is
flexible enough to fit important features in typical SN lightcurves,
but sufficiently restrictive not to allow unreasonable fitted
lightcurves, the resulting representation of the data reduces the
chances of overfitting by NN or any other machine learning
classification algorithm. Functional fitting also circumvents the
problem associated with flux measurements being made on an irregular
time grid. Details of the function fitting are given in
Sec.~\ref{sec:method:template}.

\item
The function parameters and their associated errors obtained by
fitting the analytical form to the SNe lightcurves, along with the
number of flux measurements, the maximum-likelihood value of the fit
and Bayesian evidence of the model are then used as input parameters
to a classification NN that outputs the probability that the SN under
consideration is of Type Ia. Details of NN training are
given in Sec.~\ref{sec:method:NN}.
\end{enumerate}

%============================================================
\subsection{Lightcurve fitting}\label{sec:method:template}
%============================================================
All lightcurves were fitted with the following parameterised
functional form, which is based on that used in
\citet{2009A&A...499..653B} and \citet{2010PASP..122.1415K}, but
differs in the precise parameterisation used:
\begin{equation}
f(t) = A\left[1+B(t-t_1)^2\right] \frac{e^{-(t-t_0)/\Tfall}}{1+e^{-(t-t_0)/\Trise}},
\label{eq:fitformula}
\end{equation}
where, for each SN, $t=0$ corresponds to the time of the earliest
measurement in the $r$-band lightcurve.  While this form has no
particular physical motivation, it is sufficiently general to fit the
shape of virtually all types of SNe lightcurves, including those with
double peaks (in contrast to the fitting function used by
\citealt{Newling2011}). Unlike \citet{2009A&A...499..653B}, we do not
include an additive offset parameter, since the simulated lightcurves
we will analyse have zero flux well outside the SN event by
construction. One could, of course, trivially include such a parameter
if, for example, one were analysing poorly-normalised real data for
which this is not the case. Moreover, we use a different
parameterisation for the quadratic pre-factor in $f(t)$ than that used
in \citet{2010PASP..122.1415K}, since we feel the form we adopt in
(\ref{eq:fitformula}) has a more transparent interpretation. In
particular, we restrict the parameter $B$ to be positive, since
negative values would, in general, lead to negative flux values $f(t)$
for some ranges of $t$.

A separate fit was performed in each filter for each SN. To perform
the fit, we assume the simple Gaussian likelihood function
\begin{equation}
\mathcal{L}(\mathbf{\Theta}) = 
\exp[-{\textstyle \frac{1}{2}}\chi^2(\mathbf{\Theta})],
\end{equation}
where the parameter vector $\mathbf{\Theta}=
\{A, B, t_1, t_{0}, T_{{\rm rise}}, T_{{\rm fall}}\}$ and
\begin{equation}
\chi^2 (\mathbf{\Theta})= \sum_{k=1}^{n} 
\frac{[F_k- f(t_k;\mathbf{\Theta})]^2}
{\sigma^2_k},
\label{eq:template:chisq}
\end{equation}
in which $n$ is the number of flux measurements for the supernova/filter
combination under consideration. The priors assumed on the function
parameters are listed in Table~\ref{tab:priors1}. 

\begin{figure*}
\begin{center}
\includegraphics[angle=-90,width=17.5cm]{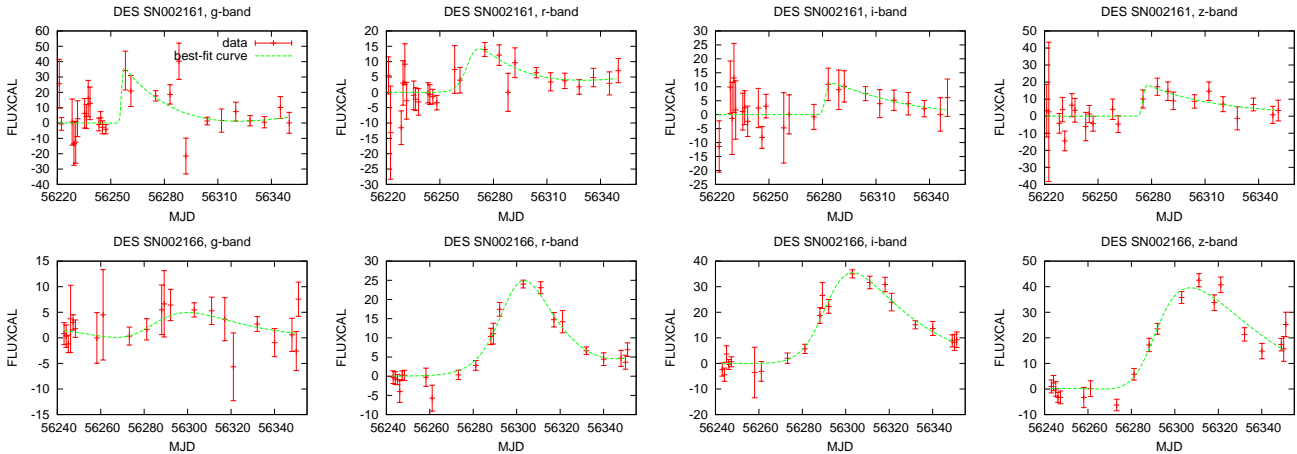}
\caption{Simulated lightcurve measurements and associated
  uncertainties (red points) in the $g$, $r$, $i$ and $z$ filters for
  a Type-Ia (top row) and non-Ia (bottom row) SN, together with the
  best-fit function (green line) of the form in equation (\ref{eq:fitformula}).}
\label{fig:fits}
\end{center}
\end{figure*}

In order to estimate the function parameters, we use the
\texttt{MultiNest} package \citep{feroz08, multinest} which is built
on the framework of nested sampling \citep{skilling04, sivia} and is
very efficient at exploring posteriors that may contain multiple modes
and/or large (curving) degeneracies, and also calculates the Bayesian
log-evidence, $\log\mathcal{Z}$, for the model. Fig.~\ref{fig:fits}
shows the data and corresponding fitted functional form in each filter
for a typical Type Ia and Type II SN, respectively. The function
$f(t)$ in (\ref{eq:fitformula}) is sufficiently flexible to provide a
good fit to all the lightcurves.

From each fit, the feature vector consisting of the mean values
$\widehat{\mathbf{\Theta}} = \{\hat{A}, \hat{B}, \hat{t}_1, \hat{t}_0,
\hat{T}_{{\rm rise},}, \hat{T}_{{\rm fall},} \}$ and standard
deviations $\bsigma = \{ \sigma_{A}, \sigma_{B}, \sigma_{t_{\rm 1}},
\sigma_{t_{\rm 0}}, \sigma_{T_{{\rm rise}}}, \sigma_{T_{{\rm fall}}}
\}$ of the one-dimensional marginalized posterior distributions of
function parameters, along with the number of flux measurements $n$,
the maximum-likelihood value and the Bayesian evidence, are then used
as inputs for NN training. For each SN, the total input vector
consists of the concatenation of the feature vectors for each filter.
Since there are 15 values in the feature vector for each filter,
resulting in a total of 60 values across all 4 filters, the function
fitting corresponds to a dimensionality reduction relative to the
number of flux measurements for some SNe, but not for others. One
facet of the robustness of our approach is that the same function
fitting process is performed for all SNe, irrespective of the number
and times relative to peak brightness of the flux measurements in each
filter. For each filter, the entire fitting process requires around 90
secs on a single CPU.

\begin{table}
\begin{center}
\begin{tabular}{ll}
\hline Lightcurve parameter & Prior \\ \hline $A$ & ${\cal U}(10^{-5},
1000)$ on $\log A$ \\ $B$ & ${\cal U}(10^{-5}, 100)$ on $\log B$
\\ $t_{1}$ & ${\cal U}(0,100)$ MJD\\ $t_{0}$ & ${\cal U}(0,100)$
MJD\\ $\Trise $ & ${\cal U}(0,100)$ MJD\\ $\Tfall $ & ${\cal
  U}(0,100)$ MJD\\ \hline
\end{tabular}
\caption{Priors on lightcurve parameters, where ${\cal U}(a,b)$ denotes a uniform distribution between the limits $a$ and $b$.\label{tab:priors1}}
\end{center}
\end{table}
%

%============================================================
\subsection{Neural network training}\label{sec:method:NN}
%============================================================

\begin{figure}
\begin{center}
\includegraphics[width=0.5\columnwidth]{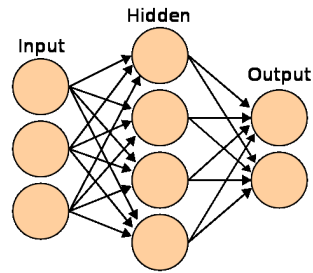}
\caption{A 3-layer neural network with 3 inputs, 4 hidden nodes, and 2 outputs. Image courtesy of Wikimedia Commons.}
\label{fig:neuralnet}
\end{center}
\end{figure}

A multilayer perceptron artificial neural network is the simplest type
of network and consists of ordered layers of perceptron nodes that
pass scalar values from one layer to the next. The perceptron is the
simplest kind of node, and maps an input vector ${\bf x} \in \Re^n$ to
a scalar output $f({\bf x};{\bf w},\theta)$ via
\begin{equation}
\label{eq:perceptron}
f({\bf x};{\bf w},\theta) = \theta + \sum_{i=1}^n {w_{i} x_{i}},
\end{equation}
where $\{w_{i}\}$ and $\theta$ are the parameters of the perceptron,
called the `weights' and `bias', respectively. We will focus mainly on
3-layer NNs, which consist of an input layer, a hidden layer, and an
output layer as shown in Fig.~\ref{fig:neuralnet}.
The outputs of the nodes in the hidden and output layers are given by the following equations:
\begin{eqnarray}
\label{eq:hiddenformula}
\textrm{hidden layer:} \; h_{j} = g^{(1)}(f^{(1)}_{j}); \; f^{(1)}_{j} = \theta^{(1)}_{j} + \sum_{l} {w^{(1)}_{jl}x_{l}} ,\\
\label{eq:outputformula}
\textrm{output layer:} \; y_{i} = g^{(2)}(f^{(2)}_{i}); \; f^{(2)}_{i} = \theta^{(2)}_{i} + \sum_{j} {w^{(2)}_{ij}h_{j}} ,
\end{eqnarray}
where $l$ runs over input nodes, $j$ runs over hidden nodes, and $i$
runs over output nodes. The functions $g^{(1)}$ and $g^{(2)}$ are
called activation functions and must be bounded, smooth, and monotonic
for our purposes. We use $g^{(1)}(x)=\tanh(x)$ and $g^{(2)}(x)=x$; the
non-linearity of $g^{(1)}$ is essential to allowing the network to
model non-linear functions.

The weights and biases are the values we wish to determine in our
training. As they vary, a huge range of non-linear mappings from
inputs to outputs is possible. In fact, a `universal approximation
theorem'~\citep{UnivApprox} states that a NN with three or more layers
can approximate any continuous function as long as the activation
function is locally bounded, piecewise continuous, and not a
polynomial.

In order to classify a set of data using a NN, we need to provide a
set of training data $ \mathcal{D} = \{{\bf x}^{(i)}, {\bf
  t}^{(i)}\}$, where ${\bf x}^{(i)}$ denotes the input parameters and
${\bf t}^{(i)}$ is a vector with membership probabilities for each
class. The likelihood function for a classification network is given
by
\begin{equation}
\mathcal{L}({\bf a}) = \sum_{i=1}^{n_{\rm t}}\sum_{j=1}^{n_{\rm c}} t_{j}^{(i)} \log 
p_{j}({\bf x}^{(i)}; {\bf a}),
\label{ClassNN-Like}
\end{equation}
where $n_{\rm t}$ is number of training data, $n_{\rm c}$ is number of
classes, $\bf a$ denotes the network weights and biases and $p_j$ is
  probability predicted by NN for class $j$. These probabilities are
  calculated using the soft-max function applied to outputs from the
  final layer of the NN:
\begin{equation}
p_{j} = \frac{e^{y_{j}}}{\sum_{j^{\prime}} e^{y_{j^{\prime}}}}.
\label{softmax}
\end{equation}

Depending on the network architecture, there can be millions of
network weights and biases which makes network training a very
complicated and computationally challenging task. Standard methods use
the gradient descent algorithm
\citep{Rumelhart:1986:LIR:104279.104293} for network training but this
does not work well for very deep networks (networks with many hidden
layers). We therefore use the \texttt{SkyNet} package for network
training which uses a $2${nd}-order optimisation method based on the
conjugate gradient algorithm and has been shown to train very deep
networks efficiently (Feroz et al., in preparation). \texttt{SkyNet}
has also been combined with \texttt{MultiNest} in the \texttt{BAMBI}
\citep{2012MNRAS.421..169G} package for fast and robust Bayesian
analysis.

\subsection{Application to post-SNPCC data}

In the original SNPCC, participants were given the
`spectroscopically-confirmed' sample $\mathcal{S}$ and asked to
predict the type of SNe in the `photometric' sample $\mathcal{P}$.  We
apply our method to this case in Sec.~\ref{sec:res}, but, as we
comment in Sec.~\ref{sec:data}, $\mathcal{S}$ is not a random subset
of $\mathcal{D}$, and so is not representative of the full data set.
As discussed in \cite{Newling2011} and \cite{2012MNRAS.419.1121R},
when training machine-learning methods -- including neural networks --
the distribution and characteristics of the training and testing
samples (see below) should be as similar as possible. It can therefore
cause difficulties to use $\mathcal{S}$ for training and $\mathcal{P}$
as the testing sample, since $\mathcal{S}$ constitutes a biased
subsample.

The original rationale for using $\mathcal{S}$ as the training data in
the SNPCC was that limited spectroscopic resources in future surveys
are likely to produce a biased sample of SNe of known
class. Nonetheless, studies of automated methods for photometric
classification of SNe may provide sufficient motivation to modify the
spectroscopic follow-up strategy \citep{2012MNRAS.419.1121R}, leading
to real spectroscopically-confirmed SNe training samples that more
closely represent the larger photometric sample.  Therefore, following
\cite{Newling2011}, after deciding on the proportions of the data to
be used for training and testing, we also consider partitioning
$\mathcal{D}$ randomly.

In fact, in each case, we divide the data among three different
categories: optimisation, validation and testing. Data in the
optimisation category constitute the $n_{\rm t}$ examples on which
network weights are optimised using the likelihood function given in
Eq.~\eqref{ClassNN-Like}. Data in the validation category are used to
guard against overfitting: when the sum of squared errors on
validation data starts to increase, the network optimisation is
stopped. The combination of optimisation and validation data, always
in the relative proportions 75:25 per cent, constitute what we call
our `training' data. Data in the testing category are not involved in
the training of the network at all, and are used to assess the
accuracy of the resulting classifier. All the results presented in
this paper are obtained from the testing data-set.

It is interesting to check the improvement in network predictions with
the amount of data used in training, we therefore construct six random
training samples $\mathcal{D}_p$ ($p=0,1,2,\ldots,5$), which contain 5, 10,
20, 30, 40 and 50 per cent of the data, respectively, and the use
remainder of the data for testing.  Sample $\mathcal{D}_0$ contains
1045 SNe and is thus similar in size to the
spectroscopically-confirmed sample $\mathcal{S}$ discussed
above. Clearly, as the amount of data used for training increases, one
should expect the accuracies of predictions coming from networks to
improve.

For each sample, we use the following input set for network
training: ${\bf x}^{(i)} = \{\hat{\mathbf{\Theta}}_{i}^{\alpha},
\hat{\mathbf{\sigma}}_{i}^{\alpha}, n_{i}^{\alpha}, (\log\mathcal
L_{\rm max})_{i}^{\alpha}, \log\mathcal{Z}_{i}^{\alpha}\}$, where $i$
indexes the SNe allocated to a given optimisation, validation and
testing category, $\alpha \in \{\rm g,r,i,z\}$ denotes the filter,
$\hat{\mathbf{\Theta}}_{i}^{\alpha}$ and $\hat{\mathbf{\sigma}}_{\rm
  i}^{\alpha}$ are the means and standard deviations respectively of
function parameters defined in Sec.~\ref{sec:method:template},
$n_{i}^{\alpha}$ is the number of flux measurements for a given SN,
$(\log\mathcal{L}_{\rm max})_{i}^{\alpha}$ is the maximum-likelihood
value of the fit and $\log\mathcal{Z}_{i}^{\alpha}$ is the Bayesian
log-evidence. Moreover, we also train further networks with redshift
$z_{i}$ and its uncertainty $\sigma_{z_{i}}$ included as additional
inputs, in order to determine whether they can improve the SNe
classification.

We use a 3-layered perceptron neural network with $500$ nodes in the
hidden layer. Training times on 16 processors for datasets ${\cal
  D}_{0}$ and ${\cal D}_5$ (i.e. the smallest and largest of those
considered), including redshift information, were $\sim 2$ and $\sim
52$~min, respectively.

%%%%%%%%%%%%%%%%%%%%%%%%%%%%%%%%%%%%%%%%%%%%%%%%%%%%%%%%%
\section{Results}\label{sec:res}
%%%%%%%%%%%%%%%%%%%%%%%%%%%%%%%%%%%%%%%%%%%%%%%%%%%%%%%%%

Once the network has been trained, it is applied to the testing
data-set to obtain the predictions for each SN therein being either
Type-Ia or non-Ia. For each SN, the NN classification requires only a
few microseconds of CPU time. To perform the classification, we first
need to pick a threshold probability $p_{\rm th}$ such that all SNe
for which the network output probability of being Type-Ia is larger
than $p_{\rm th}$ are identified as Type-Ia candidates. One can then
calculate the completeness $\epsilon_{\rm Ia}$ (fraction of all type
Ia SNe that have been correctly classified; also often called the
efficiency), purity $\tau_{\rm Ia} $(fraction of all Type Ia
candidates that have been classified correctly) and figure of merit
$\mathcal{F}_{\rm Ia}$ for Type Ia SNe. These quantities are defined
as follows:
\begin{eqnarray}
\epsilon_{\rm Ia} & = & \frac{N_{\rm Ia}^{\rm true}}{N_{\rm Ia}^{\rm total}},\\
\label{eq:completeness}
\tau_{\rm Ia} & = & \frac{N_{\rm Ia}^{\rm true}}{N_{\rm Ia}^{\rm true} + N_{\rm Ia}^{\rm false}},\\
\label{eq:purity}
\mathcal{F}_{\rm Ia} & = & \frac{1}{N_{\rm Ia}^{\rm total}} \frac{(N_{\rm Ia}^{\rm true})^2}{N_{\rm Ia}^{\rm true}+WN_{\rm Ia}^{\rm false}},
\label{eq:FoM}
\end{eqnarray}
where $N_{\rm Ia}^{\rm total}$ is the total number of Type Ia SNe in
the sample, $N_{\rm Ia}^{\rm true}$ is the number of SNe correctly
predicted to be of Type Ia, $N_{\rm Ia}^{\rm false}$ is the number of
SNe incorrectly predicted to be of Type Ia and $W$ is a penalty factor
which controls the relative penalty for false positives over false
negatives. For SNPCC $W \equiv 3$. 

\begin{table}
\begin{center}
\begin{tabular}{ccccc}
\hline
Sample & $z$ used & Completeness & Purity & FoM \\
\hline
$\mathcal{S}$ & no & 0.93 & 0.32 & 0.12 \\
$\mathcal{D}_0$ & no & 0.71 & 0.67 & 0.29 \\
$\mathcal{D}_1$ & no & 0.78 & 0.77 & 0.41 \\
$\mathcal{D}_2$ & no & 0.79 & 0.79 & 0.43 \\
$\mathcal{D}_3$ & no & 0.82 & 0.80 & 0.46 \\
$\mathcal{D}_4$ & no & 0.82 & 0.82 & 0.50 \\
$\mathcal{D}_5$ & no & 0.85 & 0.82 & 0.51 \\[1mm]
$\mathcal{S}$ & yes & 0.94 & 0.34 & 0.14 \\
$\mathcal{D}_0$ & yes & 0.75 & 0.70 & 0.33 \\
$\mathcal{D}_1$ & yes & 0.79 & 0.78 & 0.43 \\
$\mathcal{D}_2$ & yes & 0.82 & 0.81 & 0.48 \\
$\mathcal{D}_3$ & yes & 0.84 & 0.83 & 0.52 \\
$\mathcal{D}_4$ & yes & 0.84 & 0.85 & 0.55 \\
$\mathcal{D}_5$ & yes & 0.88 & 0.85 & 0.58 \\

\hline
\end{tabular}
\caption{Completeness, purity and figure of merit for Type-Ia SNe
  classification obtained by applying trained networks to the various
  testing data-sets with threshold probability $p_{\rm th}$ set to
  0.5.}
\label{tab:results}
\end{center}
\end{table}

\begin{figure*}
\begin{center}
\includegraphics[width=0.5\columnwidth, angle=-90]{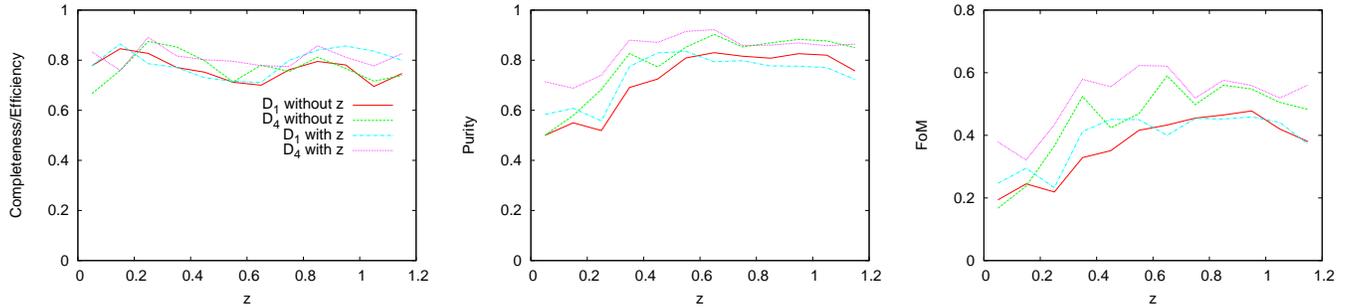}
\caption{Completeness, purity and figure of merit for Type-Ia SNe
  classification as a function of redshift ($z$) from applying trained
  networks with a threshold probability ($p_{\rm th}$) of 0.5; `with
  $z$'/`without $z$' indicates that redshift information was/was not
  used in network training. Samples $\mathcal{D}_1$ and
  $\mathcal{D}_4$ use 10 and 40 per cent of the data, respectively,
  for training.}
\label{fig:zResults}
\end{center}
\end{figure*}

Our classification results using a canonical threshold probability
$p_{\rm th} = 0.5$ are summarised in Table~\ref{tab:results}.  
 It can be clearly seen that there is a significant improvement
in the results as more training data are used. Moreover, comparing the
results for training samples $\mathcal{S}$ and $\mathcal{D}_0$, which
are similar in size, one sees that the overall quality of the
classification is much higher for the random, representative sample
$\mathcal{D}_0$ than for the biased sample $\mathcal{S}$; this agrees
with the findings of \cite{Newling2011} and
\cite{2012MNRAS.419.1121R}.  Nonetheless, although the purity obtained
for the sample $\mathcal{S}$ is quite low, the completeness is very
high. This is most likely because of the biased nature of
$\mathcal{S}$, in which about 51 per cent of SNe are Type-Ia, whereas
$\mathcal{P}$ contains only 22 per cent Type-Ia. Thus, the classifier
has not been trained with a representative collection of non-Ia SNe,
and hence often misclassifies them as Type-Ia. It is worth noting that
the `toy classifier', which simply classifies all SNe in $\mathcal{P}$
as Type Ia, would yield $\epsilon_{\rm Ia}=1.00$, $\tau_{\rm Ia}
=0.22$ and $\mathcal{F}_{\rm Ia}=0.09$.

The inclusion of redshift information results in a modest 5--10 per
cent improvement in $\epsilon_{\rm Ia}$, $\tau_{\rm Ia}$ and
$\mathcal{F}_{\rm Ia}$ for each random sample $\mathcal{D}_p$, but
only a very marginal improvement for $\mathcal{S}$.  In
Fig.~\ref{fig:zResults} we show the completeness, purity and figure of
merit for Type Ia classification as function of SNe redshift $z$ for
the training samples $\mathcal{D}_1$ and $\mathcal{D}_4$. One sees
that the completeness (efficiency) shows no dependence on $z$, but
that there is a slight drop-off in purity (and hence FoM) at low $z$.
This probably occurs because the percentage of Type Ia SN is lower at
low redshift, which would indeed be expected to reduce purity, but not
completeness. This is most easily understood by again considering the
'toy classifier' mentioned above, which would clearly display this
behaviour; we would expect facets of this generic behaviour to occur
for any classification method.

\begin{figure*}
\begin{center}
\includegraphics[width=0.5\columnwidth,
  angle=-90]{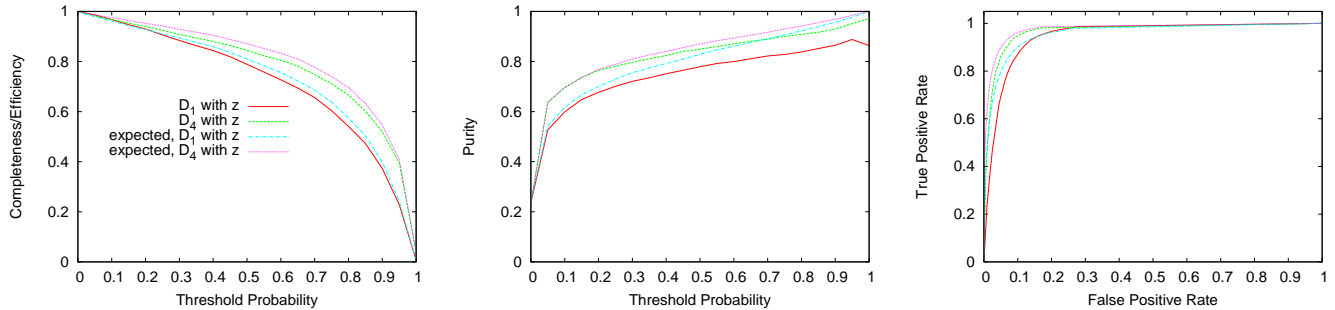}
\caption{True and expected completeness, purity and
Receiver Operating Characteristic (ROC) as a function of threshold probability
  $p_{\rm th}$ from applying trained networks to get Ia/non-Ia
  classification probabilities on the testing data-set. No redshift
  information was used in network training. Samples $\mathcal{D}_1$ and
  $\mathcal{D}_4$ use 10 and 40 per cent of the data, respectively,
  for training.}
\label{fig:expResults}
\end{center}
\end{figure*}

An important feature of probabilistic classification is that it allows
one to investigate the quality of the classification as a function of
the threshold probability. Moreover, as pointed out in
\cite{2008arXiv0810.0781F}, it also allows one to calculate the {\em
  expected} completeness, purity and figure of merit as a function of
$p_{\rm th}$, {\em without} knowing the true classes of the SNe in the
testing sample (as is the case in classification of real SNe data, in
the absence of spectroscopic follow-up observations). Equally, these
expected values can be calculated without the need to perform an
average explicitly over realisations of the testing sample.

Let us assume that the predicted probabilities for each SN being of
Type Ia are given by $p_{\rm Ia,i}$. The expected values of the total
number $\hat{N}_{\rm Ia}^{\rm total}$ of Type Ia SNe in the sample,
the number $\hat{N}_{\rm Ia}^{\rm true}$ of SNe correctly predicted to
be of Type Ia, and the number $\hat{N}_{\rm Ia}^{\rm false}$ of SNe
incorrectly predicted to be of Type Ia can then be calculated as
follows:
\begin{eqnarray}
\hat{N}_{\rm Ia}^{\rm total} & = & \sum_{i=1}^{N} p_{{\rm Ia}, i}, \\
\label{eq:ExpNTot}
\hat{N}_{\rm Ia}^{\rm true} & = & \sum_{i=1,p_{{\rm Ia}, i}>p_{\rm th}}^{N} p_{{\rm Ia}, i},\\
\label{eq:ExpNTrue}
\hat{N}_{\rm Ia}^{\rm false} & = & \sum_{i=1, p_{{\rm Ia}, i}>p_{\rm th}}^{N} 1 - p_{{\rm Ia}, i},
\label{eq:ExpNFalse}
\end{eqnarray}
where $N$ is the total number of SNe classified and $p_{\rm th}$ is
the threshold probability. We can then use
Eqs.~\eqref{eq:completeness}, \eqref{eq:purity} and \eqref{eq:FoM} to
calculated the expected values of completeness $\hat{\epsilon}_{\rm
  Ia}$, purity $\hat{\tau}_{\rm Ia}$ and figure of merit
$\hat{\mathcal{F}}_{\rm Ia}$ as a function of threshold probability
$p_{\rm th}$. 

In Fig.~\ref{fig:expResults}, we plot the actual and expected values
of completeness and purity for the samples $\mathcal{D}_1$ and
$\mathcal{D}_4$, trained without redshift information. Plots for
networks trained with redshift information are quite similar and
therefore we do not show them. One sees that the expected completeness
and purity curves match quite well with the corresponding actual
ones. Thus, in principle, rather than arbitrarily choosing the value
$p_{\rm th}=0.5$ (say), which was used to produce the results in
Table~\ref{tab:results} and Fig.~\ref{fig:zResults}, one could instead
choose a value of $p_{\rm th}$ designed to achieve a target overall
completeness and/or purity for a given survey.

We also plot the actual and expected Receiver Operating Characteristic
(ROC) curves (see e.g. \citealt{Fawcett:2006:IRA:1159473.1159475}) for
our analysis procedure in Fig.~\ref{fig:expResults}.  The ROC curve
provides a very reliable way of selecting the optimal algorithm in
signal detection theory. We employ the ROC curve here to analyse our
SNe classification criterion, based on the threshold probability
$p_{\rm th}$. The ROC curve plots the True Positive Rate (TPR) against
the False Positive Rate (FPR) as a function of the threshold
probability. TPR is, in fact, identical to completeness (and also
equals the `power' of the classification test in a Neyman--Pearson
sense); it may also be defined as the ratio of the number of true
positives for a given $p_{\rm th}$ to the number of true positives for
$p_{\rm th} = 0$. Conversely, FPR is the ratio of the number of false
positives for a given $p_{\rm th}$ to the number of false positives
for $p_{\rm th} = 0$, which is also often referred to as the
contamination (or the Neyman--Pearson type-I error rate)\footnote{It
  is worth noting that, in terms of conditional probabilities,
  completeness is simply $\Pr(\mbox{classified as Ia}|\mbox{Ia})$,
  purity is its Bayes' theorem complement
  $\Pr(\mbox{Ia}|\mbox{classified as Ia})$, and contamination is
  $\Pr(\mbox{classified as Ia}|\mbox{non-Ia})$.}.  A perfect binary
classification method would yield a ROC curve in the form of a
right-angle connecting the points $(0,0)$ and $(1,0)$ in the ROC space
via the upper left corner $(0,1)$. A completely random classifier
would yield a diagonal line connecting $(0,0)$ and $(1,0)$ directly.

One sees from the right-hand panel in Fig.~\ref{fig:expResults} that
our method yields very reasonable ROC curves, indicating that the
classifiers are quite discriminative. One also sees that our expected
ROC curves match well with the corresponding actual ones. Thus, in
principle, one could `optimise' the classifier by choosing $p_{\rm
  th}$ in a number of possible ways. For example, from among the
numerous possibilities, one could choose $p_{\rm th}$ such that it
corresponds to the point on the expected ROC curve where, either: (i)
the ROC curve crosses the straight line connecting the $(0,1)$ and
$(1,0)$ in the ROC space; or (ii) the straight line joining the point
$(1,0)$ to the ROC curve intersects it at right-angles.

%%%%%%%%%%%%%%%%%%%%%%%%%%%%%%%%%%%%%%%%%%%%%%%%%%%%%%%%%
\section{Discussion and conclusions}\label{sec:conc}
%%%%%%%%%%%%%%%%%%%%%%%%%%%%%%%%%%%%%%%%%%%%%%%%%%%%%%%%%

We have presented a new method for performing automated photometric of
SNe into Type-Ia and non-Ia. In our a two-stage approach, the SNe
lightcurves are first fitted to an analytic parameterised function,
and the resulting parameters, together with a few further
statistics associated with the fit, are then used as the input feature
vector to a classification neural network whose output is the
probability that the SN is of Type-Ia. Assuming a canonical threshold
output probability $p_{\rm th}=0.5$, when we train the method using a
random sample of 10 (40) per cent of the updated simulated data set
released following the SuperNova Photometric Classification Challenge
(post-SNPCC), making no selection cuts, we find that it yields robust
classification results, namely a completeness of 0.78 (0.82), purity
of 0.77 (0.82), and SNPCC figure-of-merit of 0.41 (0.50).  A modest
5--10 per cent improvement in these results is achieved by also
including the SN host-galaxy redshift and its uncertainty as inputs to
the classification network. The quality of the classification does not
depend strongly on the SN redshift.

It is difficult to perform a direct comparison of our results with
those submitted to the original SuperNova Photometric Classification
Challenge, which are summarised in \cite{2010PASP..122.1415K}.  As
pointed out in that paper, the original challenge data set suffered
from a number of bugs; these were subsequently corrected before the
release of the post-SNPCC data set used in this paper. The latter also
benefited from further improvement in the generation of the
simulations, leading to more realistic SNe lightcurves.  It is hard to
assess how these differences affect the difficulty of classifying the
SNe. For example, in the original SNPCC data set, non-Ia SNe were too
dim on average, which made classifcation of Type-Ia SNe easier.
Conversely, participants in the original SNPCC were given the
spectroscopically confirmed sample $\mathcal{S}$ of 1,103 SNe and
asked to predict the type of SNe in the simulated sample $\mathcal{P}$
of 20,216 SNe. As discussed in Sec.~\ref{sec:data}, the fact that
$\mathcal{S}$ is not a representative training sample makes
classification more difficult than if one simply uses random training
samples, on which most of our analysis has been focused. Despite
these caveats, for reference we note that the original challenge entry
with the highest SNPCC figure of merit of $\sim 0.4$ (averaged over SN
redshift bins), achieved an overall completeness of 0.96 and purity of
0.79, although the quality of the classification varied considerably
with SN redshift.

More recent works by \cite{Newling2011}, \cite{2012MNRAS.419.1121R}
and \cite{2012arXiv1201.6676I} analyse the same post-SNPCC data set
$\mathcal{D}$ used in this paper. A meaningful comparison of their
results with our own is still not straightforward, however, since all
three studies make different choices for the nature and size of the
subsets of $\mathcal{D}$ used for training and testing, which also
differ from the choices made in this paper. Nonetheless, some broad
comparisons are possible.  

The most straightforward comparison is with \cite{Newling2011}, who
present results using, as we do, the subset $\mathcal{S}$ and also
various random subsets of $\mathcal{D}$ (which they call
`representative samples') for training their classifier. For
$\mathcal{S}$, their KDE method achieves a figure-of-merit of 0.37
(0.39) without (with) the inclusion of host-galaxy redshift
information, whereas their boosting method yields a figure-of merit
of 0.15 using redshift information.  Turning to their analysis of
representative samples, from figure~15 in their paper, one sees that
their boosting method, which is the more successful of their two
methods on the representative samples, achieves figures-of-merit of
$\sim 0.45$ and $\sim 0.55$ for training sets containing $\sim 2000$
and $\sim 8000$ SNe, respectively, which correspond roughly to the
sizes of our $\mathcal{D}_1$ and $\mathcal{D}_4$ training data
sets. These classification results are obtained from the remaining
SNe in $\mathcal{D}$, after including SNe host-galaxy redshift
information, and are very similar to the equivalent figures-of-merit
achieved by our own classifier, as listed in
Table~\ref{tab:results}. Unfortunately, \cite{Newling2011} do not give
values for their corresponding completeness and purity values, so it
is not possible to compare their results with our own.

Both \cite{2012MNRAS.419.1121R} and \cite{2012arXiv1201.6676I} adopt
very different approaches from the above, and from each other, for
choosing the nature and size of the subsets of $\mathcal{D}$ used for
training. Indeed, each of these works considers a range of training
sets. \cite{2012MNRAS.419.1121R} do, however, consider training using
the spectroscopically-confirmed sample $\mathcal{S}$, and obtain a
purity of 0.50 (0.54), a completeness of 0.50 (0.90) and a
figure-of-merit of 0.13 (0.25) without (with) the inclusion of
redshift information. They also construct three further classes of
training sets, two of which contain several examples, each requiring
the same fixed amount of spectroscopic follow-up time assumed in the
construction of the original sample $\mathcal{S}$. These are: (i) SNe
observed in order of decreasing brightness; (ii) ($r$ band)
magnitude-limited surveys down to 23.5, 24. 24.5 and 25th magnitude,
respectively; and (iii) redshift-limited surveys out to $z=0.4$ and
$0.6$, respectively.  In applying their classifier to the post-SNPCC
the remaining part of the photometric sample $\mathcal{P}$, their best
classification results are a completeness of 0.65 (0.74), a purity of
0.72 (0.76) and a figure-of-merit of 0.31 (0.36) without (with) the
inclusion of redshift information; these were obtained from the
deepest magnitude-limited survey, which contained only 165 SNe. It is
worth noting the deeper training sets have a SNe class composition
that closely resembles that of the full data set $\mathcal{D}$, and so
begin to approximate a representative training sample.

The training sets considered by \cite{2012arXiv1201.6676I} are closer
in spirit to the one used in the original SNPCC. Starting with the
spectroscopically-confirmed subsample $\mathcal{S}$, as a requirement
of their method they impose selection cuts such that every SN must
have at least one observation epoch with $t \le t_{\rm low}$ and one
with $t \ge t_{\rm up}$ in all available filters. In addition, each SN
must have at least 3 observations above a given signal-to-noise ratio
(SNR) in each filter. The same selection cuts are also applied to the
photometric subsample $\mathcal{P}$ to produce the corresponding
testing sets. Of the selection cuts considered, their sample $D_5$
with $\mbox{SNR} > 0$ is the least restrictive, yielding training and
testing sets containing 830 and 15,988 SNe, respectively. On these
demanding data their classifier yields a completeness of 0.44, a
purity of 0.37, and a figure-of-merit of 0.06.

From these various comparisons, we conclude that the method presented
in this paper is indeed competitive, inspite of its relative
simplicity, and yields reasonably robust classifications. Finally, we
note that, aside from its relative simplicity and robustness, the
classification method that we have presented here can be extended and improved
in a number of ways. Firstly, one can easily generalise our approach to divide the non-Ia SNe into their different classes.
Moreover, our use of probabilistic
classification allows one to calculate expected completeness, purity
and figure-of-merit (amongst other measures), without knowing the true
classes of the SNe in the testing sample, as will be the case in the
classification of real SNe. This allows one to tailor the method by
adjusting the output threshold probability $p_{\rm th}$ to achieve a
given completeness, purity or figure-of-merit. Alternatively, one
could use the expected ROC curve to ‘optimise’ the value of $p_{\rm
  th}$ used for classification. Nonetheless, one must always remember
that if the training sample is not representative of the population,
then the predictions for any measure of classification quality will
inevitably be biased, in which case the derived threshold probability
may, in fact, not be optimal.  We plan to investigate these issues in a
future work.

%%%%%%%%%%%%%%%%%%%%%%%%%%%%%%%%%%%%%%%%%%%%%%%%%%%%%%%%%
\section*{Acknowledgements}\label{sec:ackn}
%%%%%%%%%%%%%%%%%%%%%%%%%%%%%%%%%%%%%%%%%%%%%%%%%%%%%%%%%
We thank Marisa March for many insightful discussions regarding
supernovae astronomy, and the two referees for their valuable suggestions.
NVK thanks Ariel Goobar for his critical, but simultaneously
motivating, comments regarding our classification method, and Johannes
Bergstrom and Chris Savage for useful comments at an early stage of
the analysis. NVK also acknowledges support from the Swedish Research
Council (contract No. 621-2010-3301).  FF is supported by a Research
Fellowship from Trinity Hall, Cambridge. This work was performed
primarily on COSMOS VIII, an SGI Altix UV1000 supercomputer, funded by
SGI/Intel, HEFCE and PPARC. The work also utilized the Darwin
Supercomputer of the University of Cambridge High Performance
Computing Service (\texttt{http://www.hpc.cam.ac.uk}), provided by
Dell Inc. using Strategic Research Infrastructure Funding from the
Higher Education Funding Council for England.

\bibliographystyle{mn2e}
\bibliography{references}

\begin{thebibliography}{}

\bibitem[\protect\citeauthoryear{{Annis}, {Cunha}, {Busha}, {Ma} \& {DES
  Collaboration}}{{Annis} et~al.}{2011}]{2011AAS...21743316A}
{Annis} J.~T.,  {Cunha} C.,  {Busha} M.,  {Ma} Z.,    {DES Collaboration} 2011,
  in American Astronomical Society Meeting Abstracts 217 Vol.~43 of Bulletin of
  the American Astronomical Society, {The Dark Energy Survey: Survey Strategy}.
p. 433.16

\bibitem[\protect\citeauthoryear{{Bazin} et~al.,}{{Bazin}
  et~al.}{2009}]{2009A&A...499..653B}
{Bazin} G.,  et~al., 2009, \aap, 499, 653

\bibitem[\protect\citeauthoryear{{Benitez-Herrera}, {R{\"o}pke}, {Hillebrandt},
  {Mignone}, {Bartelmann} \& {Weller}}{{Benitez-Herrera}
  et~al.}{2012}]{2012MNRAS.419..513B}
{Benitez-Herrera} S.,  {R{\"o}pke} F.,  {Hillebrandt} W.,  {Mignone} C.,
  {Bartelmann} M.,    {Weller} J.,  2012, \mnras, 419, 513

\bibitem[\protect\citeauthoryear{{Blake} et~al.,}{{Blake}
  et~al.}{2011}]{2011MNRAS.418.1707B}
{Blake} C.,  et~al., 2011, \mnras, 418, 1707

\bibitem[\protect\citeauthoryear{{Conley} et~al.,}{{Conley}
  et~al.}{2011}]{2011ApJS..192....1C}
{Conley} A.,  et~al., 2011, \apjs, 192, 1

\bibitem[\protect\citeauthoryear{{D'Andrea} et~al.,}{{D'Andrea}
  et~al.}{2010}]{2010ApJ...708..661D}
{D'Andrea} C.~B.,  et~al., 2010, \apj, 708, 661

\bibitem[\protect\citeauthoryear{{Dodelson} \& {Vallinotto}}{{Dodelson} \&
  {Vallinotto}}{2006}]{dodelson06}
{Dodelson} S.,  {Vallinotto} A.,  2006, \prd, 74, 063515

\bibitem[\protect\citeauthoryear{{Falck}, {Riess} \& {Hlozek}}{{Falck}
  et~al.}{2010}]{2010ApJ...723..398F}
{Falck} B.~L.,  {Riess} A.~G.,    {Hlozek} R.,  2010, \apj, 723, 398

\bibitem[\protect\citeauthoryear{Fawcett}{Fawcett}{2006}]{Fawcett:2006:IRA:1159473.1159475}
Fawcett T.,  2006, Pattern Recogn. Lett., 27, 861

\bibitem[\protect\citeauthoryear{{Feroz} \& {Hobson}}{{Feroz} \&
  {Hobson}}{2008}]{feroz08}
{Feroz} F.,  {Hobson} M.~P.,  2008, \mnras, 384, 449

\bibitem[\protect\citeauthoryear{{Feroz}, {Hobson} \& {Bridges}}{{Feroz}
  et~al.}{2009}]{multinest}
{Feroz} F.,  {Hobson} M.~P.,    {Bridges} M.,  2009, \mnras, 398, 1601

\bibitem[\protect\citeauthoryear{{Feroz}, {Marshall} \& {Hobson}}{{Feroz}
  et~al.}{2008}]{2008arXiv0810.0781F}
{Feroz} F.,  {Marshall} P.~J.,    {Hobson} M.~P.,  2008, ArXiv e-prints

\bibitem[\protect\citeauthoryear{{Fryer} et~al.,}{{Fryer}
  et~al.}{2007}]{2007PASP..119.1211F}
{Fryer} C.~L.,  et~al., 2007, \pasp, 119, 1211

\bibitem[\protect\citeauthoryear{{Gong}, {Cooray} \& {Chen}}{{Gong}
  et~al.}{2010}]{2010ApJ...709.1420G}
{Gong} Y.,  {Cooray} A.,    {Chen} X.,  2010, \apj, 709, 1420

\bibitem[\protect\citeauthoryear{{Graff}, {Feroz}, {Hobson} \&
  {Lasenby}}{{Graff} et~al.}{2012}]{2012MNRAS.421..169G}
{Graff} P.,  {Feroz} F.,  {Hobson} M.~P.,    {Lasenby} A.,  2012, \mnras, 421,
  169

\bibitem[\protect\citeauthoryear{{Hillebrandt} \& {Niemeyer}}{{Hillebrandt} \&
  {Niemeyer}}{2000}]{2000ARA&A..38..191H}
{Hillebrandt} W.,  {Niemeyer} J.~C.,  2000, \araa, 38, 191

\bibitem[\protect\citeauthoryear{{Homeier}}{{Homeier}}{2005}]{2005ApJ...620...12H}
{Homeier} N.~L.,  2005, \apj, 620, 12

\bibitem[\protect\citeauthoryear{{Hornik}, {Stinchcombe} \& {White}}{{Hornik}
  et~al.}{1990}]{UnivApprox}
{Hornik} K.,  {Stinchcombe} M.,    {White} H.,  1990, Neural Networks, 3, 359

\bibitem[\protect\citeauthoryear{{Ishida} \& {de Souza}}{{Ishida} \& {de
  Souza}}{2012}]{2012arXiv1201.6676I}
{Ishida} E.~E.~O.,  {de Souza} R.~S.,  2012, ArXiv e-prints

\bibitem[\protect\citeauthoryear{{Ivezic} et~al.,}{{Ivezic}
  et~al.}{2008}]{2008arXiv0805.2366I}
{Ivezic} Z.,  et~al., 2008, ArXiv e-prints

\bibitem[\protect\citeauthoryear{{Johnson} \& {Crotts}}{{Johnson} \&
  {Crotts}}{2006}]{2006AJ....132..756J}
{Johnson} B.~D.,  {Crotts} A.~P.~S.,  2006, \aj, 132, 756

\bibitem[\protect\citeauthoryear{{J{\"o}nsson}, {Dahl{\'e}n}, {Goobar},
  {M{\"o}rtsell} \& {Riess}}{{J{\"o}nsson} et~al.}{2007}]{jacob07}
{J{\"o}nsson} J.,  {Dahl{\'e}n} T.,  {Goobar} A.,  {M{\"o}rtsell} E.,
  {Riess} A.,  2007, \jcap, 6, 2

\bibitem[\protect\citeauthoryear{{J{\"o}nsson}, {Dahl{\'e}n}, {Hook}, {Goobar}
  \& {M{\"o}rtsell}}{{J{\"o}nsson} et~al.}{2010a}]{jonssonGOODS}
{J{\"o}nsson} J.,  {Dahl{\'e}n} T.,  {Hook} I.,  {Goobar} A.,    {M{\"o}rtsell}
  E.,  2010a, \mnras, 402, 526

\bibitem[\protect\citeauthoryear{{J{\"o}nsson}, {Sullivan}, {Hook}, {Basa},
  {Carlberg}, {Conley}, {Fouchez}, {Howell}, {Perrett} \&
  {Pritchet}}{{J{\"o}nsson} et~al.}{2010b}]{jonssonSNLS}
{J{\"o}nsson} J.,  {Sullivan} M.,  {Hook} I.,  {Basa} S.,  {Carlberg} R.,
  {Conley} A.,  {Fouchez} D.,  {Howell} D.~A.,  {Perrett} K.,    {Pritchet} C.,
   2010b, \mnras, 405, 535

\bibitem[\protect\citeauthoryear{{Karpenka}, {March}, {Feroz} \&
  {Hobson}}{{Karpenka} et~al.}{2012}]{Karpenka2012}
{Karpenka} N.~V.,  {March} M.~C.,  {Feroz} F.,    {Hobson} M.~P.,  2012, ArXiv
  e-prints

\bibitem[\protect\citeauthoryear{{Kessler} et~al.,}{{Kessler}
  et~al.}{2009}]{2009ApJS..185...32K}
{Kessler} R.,  et~al., 2009, \apjs, 185, 32

\bibitem[\protect\citeauthoryear{Kessler et~al.,}{Kessler
  et~al.}{2009}]{Kessler2009SNANA}
Kessler R.,  et~al., 2009, arXiv:0908.4280

\bibitem[\protect\citeauthoryear{{Kessler} et~al.,}{{Kessler}
  et~al.}{2010}]{2010PASP..122.1415K}
{Kessler} R.,  et~al., 2010, \pasp, 122, 1415

\bibitem[\protect\citeauthoryear{{Kronborg} et~al.,}{{Kronborg}
  et~al.}{2010}]{kronborg10}
{Kronborg} T.,  et~al., 2010, \aap, 514, A44

\bibitem[\protect\citeauthoryear{{Kunz}, {Bassett} \& {Hlozek}}{{Kunz}
  et~al.}{2007}]{2007PhRvD..75j3508K}
{Kunz} M.,  {Bassett} B.~A.,    {Hlozek} R.~A.,  2007, \prd, 75, 103508

\bibitem[\protect\citeauthoryear{{Kuznetsova} \& {Connolly}}{{Kuznetsova} \&
  {Connolly}}{2007}]{2007ApJ...659..530K}
{Kuznetsova} N.~V.,  {Connolly} B.~M.,  2007, \apj, 659, 530

\bibitem[\protect\citeauthoryear{{Mantz}, {Allen}, {Rapetti} \&
  {Ebeling}}{{Mantz} et~al.}{2010}]{2010MNRAS.406.1759M}
{Mantz} A.,  {Allen} S.~W.,  {Rapetti} D.,    {Ebeling} H.,  2010, \mnras, 406,
  1759

\bibitem[\protect\citeauthoryear{{March}, {Trotta}, {Berkes}, {Starkman} \&
  {Vaudrevange}}{{March} et~al.}{2011}]{march11}
{March} M.~C.,  {Trotta} R.,  {Berkes} P.,  {Starkman} G.~D.,    {Vaudrevange}
  P.~M.,  2011, \mnras, 418, 2308

\bibitem[\protect\citeauthoryear{{Metcalf}}{{Metcalf}}{1999}]{metcalf99}
{Metcalf} R.~B.,  1999, \mnras, 305, 746

\bibitem[\protect\citeauthoryear{{Metcalf} \& {Silk}}{{Metcalf} \&
  {Silk}}{1999}]{metcSilk}
{Metcalf} R.~B.,  {Silk} J.,  1999, \apjl, 519, L1

\bibitem[\protect\citeauthoryear{{Newling}, {Varughese}, {Bassett}, {Campbell},
  {Hlozek}, {Kunz}, {Lampeitl}, {Martin}, {Nichol}, {Parkinson} \&
  {Smith}}{{Newling} et~al.}{2011}]{Newling2011}
{Newling} J.,  {Varughese} M.,  {Bassett} B.,  {Campbell} H.,  {Hlozek} R.,
  {Kunz} M.,  {Lampeitl} H.,  {Martin} B.,  {Nichol} R.,  {Parkinson} D.,
  {Smith} M.,  2011, \mnras, 414, 1987

\bibitem[\protect\citeauthoryear{{Nugent}, {Kim} \& {Perlmutter}}{{Nugent}
  et~al.}{2002}]{2002PASP..114..803N}
{Nugent} P.,  {Kim} A.,    {Perlmutter} S.,  2002, \pasp, 114, 803

\bibitem[\protect\citeauthoryear{{Perlmutter} et~al.,}{{Perlmutter}
  et~al.}{1999}]{1999ApJ...517..565P}
{Perlmutter} S.,  et~al., 1999, \apj, 517, 565

\bibitem[\protect\citeauthoryear{{Poznanski}, {Gal-Yam}, {Maoz}, {Filippenko},
  {Leonard} \& {Matheson}}{{Poznanski} et~al.}{2002}]{2002PASP..114..833P}
{Poznanski} D.,  {Gal-Yam} A.,  {Maoz} D.,  {Filippenko} A.~V.,  {Leonard}
  D.~C.,    {Matheson} T.,  2002, \pasp, 114, 833

\bibitem[\protect\citeauthoryear{{Poznanski}, {Maoz} \& {Gal-Yam}}{{Poznanski}
  et~al.}{2007}]{2007AJ....134.1285P}
{Poznanski} D.,  {Maoz} D.,    {Gal-Yam} A.,  2007, \aj, 134, 1285

\bibitem[\protect\citeauthoryear{{Rauch}}{{Rauch}}{1991}]{rauch91}
{Rauch} K.~P.,  1991, \apj, 374, 83

\bibitem[\protect\citeauthoryear{{Richards}, {Homrighausen}, {Freeman},
  {Schafer} \& {Poznanski}}{{Richards} et~al.}{2012}]{2012MNRAS.419.1121R}
{Richards} J.~W.,  {Homrighausen} D.,  {Freeman} P.~E.,  {Schafer} C.~M.,
  {Poznanski} D.,  2012, \mnras, 419, 1121

\bibitem[\protect\citeauthoryear{{Riess} et~al.,}{{Riess}
  et~al.}{1998}]{1998AJ....116.1009R}
{Riess} A.~G.,  et~al., 1998, \aj, 116, 1009

\bibitem[\protect\citeauthoryear{{Rodney} \& {Tonry}}{{Rodney} \&
  {Tonry}}{2009}]{2009ApJ...707.1064R}
{Rodney} S.~A.,  {Tonry} J.~L.,  2009, \apj, 707, 1064

\bibitem[\protect\citeauthoryear{Rumelhart, Hinton \& Williams}{Rumelhart
  et~al.}{1986}]{Rumelhart:1986:LIR:104279.104293}
Rumelhart D.~E.,  Hinton G.~E.,    Williams R.~J.,  1986, MIT Press, Cambridge,
  MA, USA, Chapt. Learning internal representations by error propagation, pp
  318--362

\bibitem[\protect\citeauthoryear{{Sako}, {Bassett}, {Connolly}, {Dilday},
  {Cambell}, {Frieman}, {Gladney}, {Kessler}, {Lampeitl}, {Marriner}, {Miquel},
  {Nichol}, {Schneider}, {Smith} \& {Sollerman}}{{Sako}
  et~al.}{2011}]{2011ApJ...738..162S}
{Sako} M.,  {Bassett} B.,  {Connolly} B.,  {Dilday} B.,  {Cambell} H.,
  {Frieman} J.~A.,  {Gladney} L.,  {Kessler} R.,  {Lampeitl} H.,  {Marriner}
  J.,  {Miquel} R.,  {Nichol} R.~C.,  {Schneider} D.~P.,  {Smith} M.,
  {Sollerman} J.,  2011, \apj, 738, 162

\bibitem[\protect\citeauthoryear{{Sako} et~al.,}{{Sako}
  et~al.}{2008}]{2008AJ....135..348S}
{Sako} M.,  et~al., 2008, \aj, 135, 348

\bibitem[\protect\citeauthoryear{{Schmidt}, {Keller}, {Francis} \&
  {Bessell}}{{Schmidt} et~al.}{2005}]{2005AAS...206.1509S}
{Schmidt} B.~P.,  {Keller} S.~C.,  {Francis} P.~J.,    {Bessell} M.~S.,  2005,
  in American Astronomical Society Meeting Abstracts \#206 Vol.~37 of Bulletin
  of the American Astronomical Society, {The SkyMapper Telescope and Southern
  Sky Survey}.
p.~457

\bibitem[\protect\citeauthoryear{Sivia \& Skilling}{Sivia \&
  Skilling}{2006}]{sivia}
Sivia D.,  Skilling J.,  2006, Data Analysis A Bayesian Tutorial.
Oxford University Press

\bibitem[\protect\citeauthoryear{{Skilling}}{{Skilling}}{2004}]{skilling04}
{Skilling} J.,  2004, in {Fischer} R.,  {Preuss} R.,   {Toussaint} U.~V.,  eds,
  American Institute of Physics Conference Series Vol.~119, {Nested Sampling}.
pp 1211--1232

\bibitem[\protect\citeauthoryear{{Sullivan} et~al.,}{{Sullivan}
  et~al.}{2006}]{2006AJ....131..960S}
{Sullivan} M.,  et~al., 2006, \aj, 131, 960

\bibitem[\protect\citeauthoryear{{Sullivan} et~al.,}{{Sullivan}
  et~al.}{2011}]{2011ApJ...737..102S}
{Sullivan} M.,  et~al., 2011, \apj, 737, 102

\bibitem[\protect\citeauthoryear{{Tyson}}{{Tyson}}{2002}]{2002SPIE.4836...10T}
{Tyson} J.~A.,  2002, in {Tyson} J.~A.,  {Wolff} S.,  eds, Society of
  Photo-Optical Instrumentation Engineers (SPIE) Conference Series Vol.~4836 of
  Society of Photo-Optical Instrumentation Engineers (SPIE) Conference Series,
  {Large Synoptic Survey Telescope: Overview}.
pp 10--20

\bibitem[\protect\citeauthoryear{{Wester} \& {Dark Energy Survey
  Collaboration}}{{Wester} \& {Dark Energy Survey
  Collaboration}}{2005}]{2005ASPC..339..152W}
{Wester} W.,  {Dark Energy Survey Collaboration} 2005, in {Wolff} S.~C.,
  {Lauer} T.~R.,  eds, Observing Dark Energy Vol.~339 of Astronomical Society
  of the Pacific Conference Series, {Dark Energy Survey and Camera}.
p.~152

\bibitem[\protect\citeauthoryear{{Zentner} \& {Bhattacharya}}{{Zentner} \&
  {Bhattacharya}}{2009}]{zentner09}
{Zentner} A.~R.,  {Bhattacharya} S.,  2009, \apj, 693, 1543

\end{thebibliography}

\label{lastpage}
\end{document}